\documentclass[times,authoryear,5p]{elsarticle}
\usepackage{graphicx,amssymb,color}

\journal{Icarus}
\begin{document}
\begin{frontmatter}
\title{Oort Cloud and Scattered Disc formation during a late dynamical instability in the Solar System}
\author{R. Brasser\corref{ca}}
\cortext[ca]{Corresponding author}
\ead{brasser\_astro@yahoo.com}
\address{Institute of Astronomy and Astrophysics, Academia Sinica, P. O. Box 23-141, Taipei 106, Taiwan}
\author{A. Morbidelli}
\ead{morby@oca.eu}
\address{Departement Lagrange, University of Nice - Sophia Antipolis, CNRS, Observatoire de la C\^{o}te d'Azur; Nice, France}
\begin{abstract}
One of the outstanding problems of the dynamical evolution of the outer solar system concerns the observed population ratio between the
Oort Cloud (OC) and the Scattered Disc (SD): observations suggest that this ratio lies between 100 and 1\,000 but simulations that
produce these two reservoirs simultaneously consistently yield a value of the order of 10. Here we stress that the populations in the
OC and SD are inferred from the observed fluxes of new Long Period Comets (LPCs) and Jupiter-family comets (JFCs), brighter than some
reference total magnitude. However, the population ratio estimated in the simulations of formation of the SD and OC refers to objects
bigger than a given size. There are multiple indications that LPCs are intrinsically brighter than JFCs, i.e. an LPC is smaller than a
JFC with the same total absolute magnitude. When taking this into account we revise the SD/JFC population ratio from our simulations
relative to Duncan and Levison (1997), and then deduce from the observations that the size-limited population ratio between the OC
and the SD is 44$_{-34}^{+54}$. This is roughly a factor of four higher than the value 12 $\pm$ 1 that we obtain in simulations where
the OC and the SD form simultaneously while the planets evolve according to the so-called `Nice model'. Thus, we still have a
discrepancy between model and `observations’, but the agreement cannot be rejected by the null hypothesis. 
\end{abstract}
\end{frontmatter}
\section{Introduction and background}
When examining the orbital data of new comets entering the inner solar system, Oort (1950) discovered that the distribution of
reciprocal semi-major axis of these comets showed a distinct excess for $1/a<5 \times 10^{-4}$~AU$^{-1}$. The observed semi-major axis
distribution led Oort to suggest that the Sun is surrounded by a cloud of comets in the region between 20\,000~AU to 150\,000~AU,
and that it contains approximately $10^{11}$ comets with isotropic inclination and random perihelia. This hypothesised cloud of comets
surrounding the Sun is now called the 'Oort cloud' (OC). \\

The formation and evolution of this reservoir of comets has been an issue of study ever since its discovery. The main uncertainties
are its population, how it formed and how its existence ties in with what we know about the evolution of the rest of the outer solar
system. We review each of these below and then state the motive behind this study.

\subsection{Total population of the Oort cloud}
The only method with which we can infer the number of comets in the OC is by determining the flux of long-period comets (LPCs), which
can be divided into new comets (NCs) and returning comets (RCs). The NCs are a proxy for the total population of the OC: the total
population of the cloud can be inferred from their flux through the inner solar system (Wiegert \& Tremaine, 1999). New comets are
traditionally classified as those with semi-major axis $a>10$~kAU (e.g. Wiegert \& Tremaine, 1999).\\

There exist two agents which perturb the comets in the cloud onto orbits that enter the inner solar system: passing stars (Weissman,
1980; Hills, 1981) and the Galactic tide (Heisler \& Tremaine, 1986; Levison et al., 2001). The passing stars cause usually small
random deviations in the orbital energy and other orbital elements. The Galactic tide on the other hand systematically modifies the
angular momentum of the comets at constant orbital energy. Heisler \& Tremaine (1986) discovered that if the semi-major axis of the
comet is large enough then the comet's change in perihelion can exceed 10~AU in a single orbit and 'jump' across the orbits of Jupiter
and Saturn and thus not suffer their perturbations before becoming visible; in this particular case the comet is considered an NC.
However, Heisler \& Tremaine's (1986) approximation of the Galactic tide only used the vertical component, which is an order of
magnitude stronger than the radial components. In this approximation the $z$-component of the comet's orbital angular momentum in the
Galactic plane is conserved and the comets follow closed trajectories in the $q-\omega$ plane (with $q$ being the perihelion distance
and $\omega$ being the argument of perihelion). Including the radial tides breaks this conservation and the flux of comets to the inner
solar system from the OC is increased (Levison et al., 2006). The trajectories that lead comets into the inner solar system should be
quickly depleted, were it not for the passing stars to refill them (Rickman et al, 2008). The synergy between these two perturbing
agents ensures there is a roughly steady supply of comets entering the inner solar system.\\

Even though the perturbations on the cloud are now understood, there remain large uncertainties in the total number of comets in the
cloud and there have been many attempts to constrain it (e.g. Oort, 1950; Hills, 1981; Weissman, 1983, 1996; Weigert \& Tremaine, 1999;
Francis, 2005). The general consensus seems the total number is between $10^{11}$ to $10^{12}$ comets with total absolute magnitude
$H_T\leq11$. Here $H_T$ is given by
\begin{equation}
H_T=V_{\rm{dis}}-5\log {\mathcal{D}} - 2.5\nu\log r
\end{equation}
where $V_{\rm{dis}}$ is the apparent magnitude (nucleus with coma), ${\mathcal{D}}$ is the distance of the comet to Earth, $\nu$ is a
measure of how the brightness scales with heliocentric distance (the photometric index), and $r$ is the distance of the comet to the
Sun. Thus, $H_T$ is the magnitude of a comet (nucleus plus coma) if viewed from the Sun, at a distance of 1~AU. There is a great
variability in $\nu$ among comets. The first to catalogue this quantity for a large sample of comets was Whipple (1978), who found that
for NCs on their inbound leg $\langle \nu \rangle = 2.44\pm 0.3$ and on their outbound leg $\langle \nu \rangle = 3.35\pm 0.4$. For
short-period comets $\nu$ is usually higher than 3. From a limited sample of LPCs Sosa \& Fern\'{a}ndez (2011) find $\langle \nu
\rangle \sim 3$. Using $\nu = 4$ yields to the commonly-used value $H_{10}$, which is close to Whipple's (1978) average for both LPCs
and short-period comets. A value $H_T \leq 11$ is used throughout the literature when referring to the total number of comets in the
OC, and thus we do so as well.\\

The best estimate of the number of comets in the cloud comes from a dynamical study by Kaib \& Quinn (2009). Based on a suggestion by
Levison et al. (2001), Kaib \& Quinn (2009) find that the fraction per unit time of the OC that is visible in the form of a new comet
is 10$^{-11}$~yr$^{-1}$. We shall adopt this fraction at a later stage in this paper. By calibrating this probability to the flux of
new comets from Francis (2005) Kaib \& Quinn (2009) conclude that the whole OC contains (2--3)$\times 10^{11}$ comets. Thus the most
recent simulations, and observations, suggest that the cloud contains approximately $(1-5) \times 10^{11}$ comets, rather than
10$^{12}$, of size equivalent to that of an LPCs with $H_T<11$. 

\subsection{Formation and evolution}
The first attempt to form the OC by direct numerical integration was undertaken by Duncan et al. (1987), who found that comets with
original perihelion $q \gtrsim 15$~AU were likely to reach the cloud while those with shorter perihelion distance were not. Duncan et
al. (1987) found that the inner edge of the cloud is located at approximately 3\,000~AU while the (assumed) outer edge is at
200\,000~AU. However, some aspects of their results, in particular their high formation efficiency, are questionable because their
initial conditions turn out not to be representative of the reality: their assumption that the first stage of scattering does not
greatly change the perihelion distance is incorrect.\\

Dones et al. (2004) performed a study similar to Duncan et al. (1987) but with more realistic initial conditions. At the end of the
simulation Dones et al. (2004) obtained a formation efficiency of only 5\%. Similar results are reported elsewhere (e.g. Kaib \& Quinn,
2008; Brasser et al., 2010).\\

However, simulations of OC formation have suffered from a difficult problem: they are unable to reproduce the inferred ratio between
the population of the Scattered Disc (SD) (Duncan \& Levison, 1997) and the OC. The SD is believed to be the source of the
Jupiter-family comets (JFCs) (Duncan \& Levison, 1997). The JFCs are a set of comets whose Tisserand parameter with respect to Jupiter
satisfies $T_J \in (2,3]$ (Levison, 1996). Care has to be taken here because this criterion would erroneously classify some
comets as JFCs when they have $T_J<3$ due to i) a high inclination but are entirely outside of Jupiter's orbit, or ii) very low
inclination objects that have their perihelia close to Jupiter but very long semi-major axis (Gladman et al., 2008). Thus we follow
Gladman et al. (2008) and additionally impose that a JFC must have $q<7.35$~AU so that the object is most likely dynamically controlled
by Jupiter.\\ 

From the current population of JFCs, Duncan \& Levison (1997) and Levison et al. (2008a) estimate there are $6 \times 10^8$ bodies in
the SD whose size is equivalent to that of JFCs with $H_T<9$. This estimate of the SD population is approximately a factor of 150 to
750 lower than the estimate of the OC population. Compensating for the difference in $H_T$ used in the derivations for the OC
($H_T<11$) and SD ($H_T<9$) populations is difficult because JFCs and LPCs appear to have strongly different $H$-distributions in the
9-11 range (Fern\'{a}ndez et al., 1999; Fern\'{a}ndez \& Sosa, 2012; Francis, 2005). Numerical simulations tend to yield a 5:1 (Kaib \&
Quinn, 2008) to 20:1 ratio (Dones et al, 2004; Brasser et al., 2010). Thus, simulations consistently underestimate the observed OC to
SD population ratio by at least an order of magnitude! This large discrepancy forms the motivation of our study.\\

Two solutions have been proposed to remedy this problem: forming part of the OC while the Sun was still in its birth cluster
(Fern\'{a}ndez \& Brun\'{i}ni, 2000; Brasser et al., 2006, 2012; Kaib \& Quinn, 2008), or forming the OC by the capture of comets from
other stars (Levison et al., 2010). Even though these scenaria appear to be able to solve the SD to OC population discrepancy,
unfortunately both could suffer from the difficulty of scattering small comets to large distances in the presence of gas drag
(Brasser et al., 2007). In light of the above problems, we decided to re-examine the whole problem from scratch, as detailed below.

\subsection{Our approach}
Scattering small comets to large distances in the presence of gas is very difficult. Therefore we suggest that the SD and the OC formed
together after the removal of the gas in the proto-planetary disc. The Nice model provides the natural framework for such a `late'
(relative to gas removal) and contemporary formation of both these reservoirs. The Nice model argues that, during the gas-disc phase,
the giant planets had orbits more circular and were more closely packed than now. The current planetary orbits have been achieved
during a phase of dynamical instability of the planets that occurred after gas removal (Tsiganis et al., 2005; Morbidelli et al.,
2007), possibly as late as the Late Heavy Bombardment event (approximately 500~Myr after gas removal; Gomes et al., 2005; Levison et
al., 2011). During this planetary instability, a primordial trans-Neptunian disc of planetesimals was dispersed, with just a few of its
objects surviving today in the Kuiper Belt, in the SD, in the OC and in the Trojan populations of Jupiter and Neptune (Morbidelli et
al., 2005; Levison et al. 2008a,b). The previous simulations addressing the OC/SD ratio (e.g. Dones et al. 2004,; Kaib \& Quinn,
2008; Brasser et al., 2010) assumed that the planets were on their current orbits, which is an unlikely scenario. Thus, in section 4.1
we re-asses this ratio in the framework of the Nice model, where the different orbital evolution of the planets might in principle lead
to a different result.\\

The higher eccentricities and inclinations of Uranus and Neptune during their migration produces a dynamically hot SD, where the
inclination distribution ranges up to several tens of degrees, consistent with the observed distribution of large SD objects. This SD
has a very different structure from that used in Duncan \& Levison (1997), which was dynamically cold (inclinations up to approximately
15$^\circ$). It is widely believed that the JFCs originate in the SD and thus the two should be intimately linked (Duncan \& Levison,
1997). In principle, the population ratio between the SD and JFCs after giant planet migration may be different from that estimated in
Duncan \& Levison (1997). We investigate this in section 4.2, leading to a new estimate of the SD population from the observed JFC
population. The OC produced in the Nice model is equivalent to that formed in previous studies (e.g. Dones et al., 2004; Kaib \& Quinn,
2008; Brasser et al., 2010), and therefore we do not need to re-evaluate the relationship between the OC population and the LPC flux.\\

Finally, in section 5 we re-examine the LPC and JFC populations. We take into account that there are multiple indications that LPCs are
intrinsically brighter than JFCs: an LPC of size comparable to that of a JFC with $H_T \sim 9$ has $H_T \sim 6.5$. Together with
the results of section 4.2 we re-assess the `observed' OC/SD ratio for a size-limited population. These are then compared with the
results obtained from the Nice-model simulation from section 4.1. Finally, we draw our conclusions in section 6.

\section{Planetary evolution}
\begin{figure}
\resizebox{\hsize}{!}{\includegraphics[angle=-90]{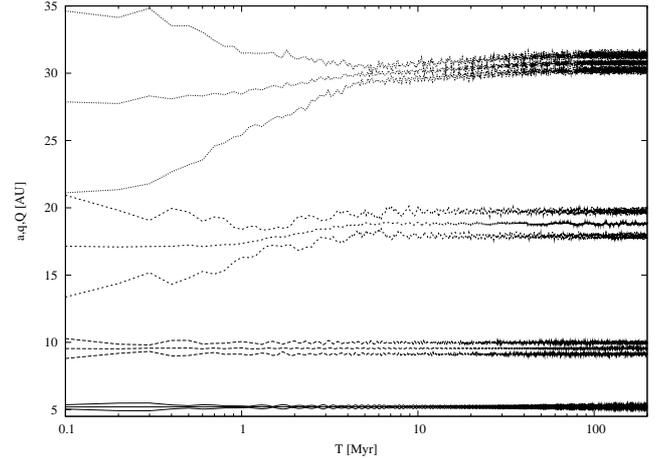}}
\caption{Evolution of the planets. The curves correspond to semi-major axis, pericentre and apocentre of the planets. Jupiter stays
close to 5~AU, Saturn close to 9~AU, Uranus ends up close to 19~AU and Neptune settles at 31~AU.}
\label{plevo}
\end{figure}

For the purpose of this study we have used one of the Nice-model planetary evolutionary tracks presented in Levison et al. (2008b).
More precisely, using the recipe described in Levison et al. (2008b), we re-enact the evolution shown in their Run A in which, at the
end of the phase of mutual close encounters among the planets, Neptune’s semi-major axis is $a_N = 27.5$~AU and its eccentricity is
$e_N = 0.3$. Uranus’s semi-major axis and eccentricity are $a_U = 17.5$~AU and $e_U = 0.2$. The mutual inclination of both planets is
approximately 1 degree. The evolution of the planets after the mutual encounters is depicted in Fig.~\ref{plevo}, where we plot the
semi-major axis, perihelion distance and aphelion ($Q$) distance for the four giant planets as a function of time. Jupiter remains
at 5~AU, Saturn stays at 9.5~AU. Uranus migrates from 17~AU to nearly 19~AU while Neptune migrates outwards to settle close to 31~AU.\\

We want to emphasize that the real evolution of the planets cannot be traced, so that we do not expect that the evolution we consider
is exactly right. However, the evolution of Run A leads to final planetary orbits very similar to the current ones and shows a high
compatibility with the currently known orbital structure of the Kuiper Belt (Levison et al., 2008b). Hence we argue the evolution
above is representative of what could have happened in reality.\\

Having decided on the evolution of the planets we explain our numerical methods in the next section.

\section{Numerical methods}
In this section we describe our numerical methods. We ran two series of simulations: one set for the formation of the OC and
another set for the formation and evolution of the SD. The planetary evolution for both sets of simulations is the same. The reason
we examine the SD separately is because we need a high resolution (large number of comets) to determine the JFC to SDO population
ratio. This high number of comets in the SD simulations was achieved by repeated cloning of the remaining comets at several stages. In
addition, for the evolution of the SD, in particular for SDOs to become JFCs, we do not need the influence of the Galactic tide and
passing stars.\\

\subsection{Oort cloud}
The simulations that were performed for the formation of the OC consist of two stages. During the first stage, the planetary evolution
shown in Fig.~\ref{plevo} is re-enacted for a total duration of 200~Myr.
We added 6\,000 massless test particles to each simulation.
Their initial conditions were taken from Levison et al. (2008b): the semi-major axes were between 29~AU and 34~AU, their eccentricities
were 0.15 and their orbits were coplanar. The time-step for the simulations was 0.4~yr. The perturbations from the Galactic tide were
included using the method of Levison et al. (2001), which incorporates both the vertical and radial Galactic tides. The local Galactic
density was 0.1~M$_{\odot}$ pc$^{-3}$ (Holmberg \& Flynn, 2000). We assumed a flat Galactic rotation curve with the Sun having an
angular velocity of 30.5 km s$^{-1}$ kpc$^{-1}$ (MacMillan \& Binney, 2010). Passing stars were included according to the method
described in Rickman et al. (2008) at a Galactic distance of 8~kpc, with the Sun's sphere of influence being 1~pc. We ran a total of
five realisations. The difference between each simulation lies in the different initial conditions for the test particles.\\

After the first stage was completed, we took the positions and velocities of the planets and test particles and resumed the integration
for another 4~Gyr using SCATR (Kaib et al., 2011), which is a Symplectically-Corrected Adaptive Timestepping Routine. It is based on
SWIFT's RMVS3 (Levison \& Duncan, 1994). It has a speed advantage over SWIFT's RMVS3 or MERCURY (Chambers, 1999) for objects far away
from both the Sun and the planets where the time step is increased. We set the boundary between the regions with short and
long time step at 300~AU from the Sun (Kaib et al. 2011). Closer than this distance the computations are performed in the heliocentric
frame, like SWIFT's RMVS3, with a time step of 0.4~yr. Farther than 300~AU, the calculations are performed in the barycentric frame
and we increased the time step to 50~yr. The error in the energy and angular momentum that is incurred every time an object crosses the
boundary at 300~AU is significantly reduced through the use of symplectic correctors (Wisdom et al., 1996). For the parameters we
consider, the cumulative error in energy and angular momentum incurred over the age of the solar system is of the same order or smaller
than that of SWIFT's RMVS3. The same Galactic and stellar parameters as in the first simulation were used. Comets were removed once
they were further than 1~pc from the Sun, or collided with the Sun or a planet.\\

\subsection{Scattered Disc}
For the simulations of the SD we used the following strategy. We took the planets and comets at the end of the planetary migration
phase and removed all comets that were further than 3\,000~AU from the Sun. Unlike the above case for the OC, to correctly
simulate the decay of the SD the planets need to be on their current orbits, or match these as closely as possible. Therefore after the
first stage of migration was completed, we performed a second stage where we artificially migrated Uranus outwards by 0.25~AU to its
current orbit over a time scale of 5~Myr. We kept Neptune's semi-major axis at its final position of 30.7~AU because lowering it to its
current value of 30.1~AU would cause resonant objects in the SD to escape from their resonances. We also had to damp the eccentricities
of Uranus and Neptune somewhat. Neptune's eccentricity was lowered from 0.015 to 0.01 while Uranus' eccentricity was lowered from
0.06 to its current value of 0.04. However, it matches its current secular properties and at its secular maximum it is 0.06. The
migration was accomplished by interpolation of the planets' orbital elements with SWIFT RMVS3 (Petit et al., 2001; Brasser et al.,
2009; Morbidelli et al., 2010). During this migration we kept the precession frequencies of the planets as close as possible to their
current values. At the end of this fictitious migration and eccentricity damping the giant planets had their current secular
architecture. During this short simulation comets were removed once they hit a planet, hit the Sun or were farther than 3\,000~AU from
the Sun. The passing stars and the Galactic tide were not included. \\

After Uranus was artificially migrated and the planets resided on their correct orbits, we proceeded to integrate the planets and
comets for another 3.8~Gyr using SWIFT RMVS3. We cloned all comets that remained after the migration three times. Cloning was
achieved by adding a random deviation of $10^{-6}$ radians to the comets' mean anomaly, keeping all the other elements fixed. We
stopped the simulations at 1~Gyr and 3.5~Gyr to clone the remaining comets three times. This repeated cloning ensured enough
comets remained during the last 500~Myr for good statistics on JFC production. \\

During the last 500~Myr it was essential to keep track of visible JFC production since we shall use this population as a proxy
for the number of comets in the Scattered Disc (Duncan \& Levison, 1997). Here we copy the term `visible JFC' from Levison \&
Duncan (1997) to refer to a JFC with perihelion distance $q<2.5$~AU. We modified SWIFT RMVS3 to output all particles that have
$q<2.5$~AU every 100~yr. Afterwards we filter out the visible JFCs by requiring they obey $q<2.5$~AU and $T_J \in [2,3]$. Here $T_J$
is the Tisserand parameter of the comet with respect to Jupiter. We also ran the last 500~Myr in a second set of simulations where we
emulated Levison \& Duncan (1997) and removed any particle the moment it came closer than 2.5~AU from the Sun. These last simulations
were done to accurately determine the fraction of SDOs that become visible JFCS, $f_{\rm{vJFC}}$.\\

All simulations were performed on either TIARA Grid or the ASIAA Condor pool. For the OC each simulation lasted only about a week,
thanks to SCATR's speed. For the SD simulations, reaching 4~Gyr took a couple of months of computation time per simulation.

\section{Results from numerical simulations}
\subsection{Formation of the Oort Cloud and Scattered Disc and prediction of the OC/SD ratio}
There have been many several publications that have studied OC formation in the current galactic environment (Dones et al., 2004; Kaib
\& Quinn, 2008; Dybczy\'{n}ski et al., 2008; Neslu\v{s}an et al., 2009; Leto et al., 2009; Brasser et al., 2010) and thus we choose not
to do an in-depth analysis of the structure of the cloud. Instead we list a few key issues because most properties of the cloud
produced in the Nice model are unlikely to be very different from previous works.\\

We define a comet to be in the OC when it has both $a >1\,000$~AU and $q>40$~AU because when $a\sim 2\,000$~AU the Galactic tide
begins to dominate over planetary perturbations from Neptune. The condition $q>40$~AU is imposed to ensure that the object has been
decoupled from Neptune by the Galactic tide. A SD object is defined to have $a<1\,000$~AU and $q > 30$~AU because its motion is
controlled by Neptune. For the sake of completeness, an object with $q \in [5,30]$~AU and not in resonance with Neptune is considered a
Centaur and an object with $q \in [30,40]$ and $a>1\,000$~AU is called a high-$a$ SDO (HaSDO) even though these could be low-$q$
OC objects. For all three populations no restrictions were placed on the inclination. We realise that our classification is not
complete because it leaves out resonant Neptune-crossing objects like Pluto, but since these only comprise a very small subset of all
objects we believe our classification is justified. Thus, we restrict ourselves to $a<1\,000$~AU for SD objects (SDOs).\\

\begin{figure}
\resizebox{\hsize}{!}{\includegraphics[angle=-90]{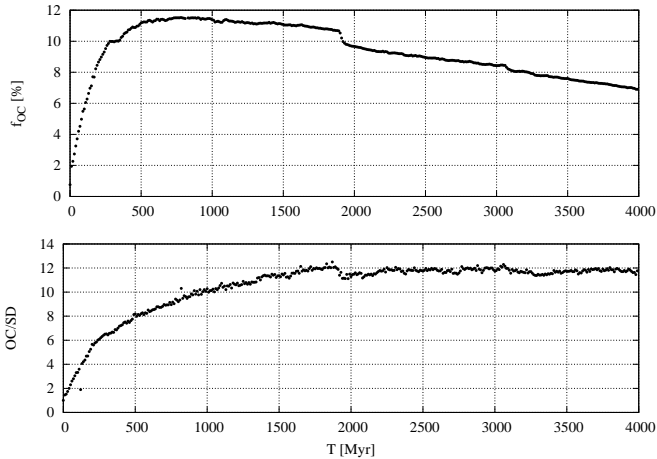}}
\caption{Top panel: Percentage of comets trapped in the OC. The bottom panel depicts the population ratio between the OC and the SD.}
\label{ocsd}
\end{figure}

In Fig.~\ref{ocsd} we have plotted the percentage of the original comets that are part of the OC (top panel) and the population ratio
between the OC and the SD (bottom panel) for one simulation. The results from the other simulations are very similar. We find that the
average OC formation efficiency at 4~Gyr is 7.1\% $\pm$ 0.3\% (1-sigma error, used throughout this paper). This value is higher than
that reported elsewhere in the literature (e.g. Dones et al. 2004; Kaib \& Quinn 2008; Dybczy\'{n}ski et al. 2008; Brasser et al.,
2010). The reason for the higher efficiency is twofold. First, all of our comets were initially placed in the vicinity of Neptune.
Second, Uranus and Neptune were initially on more eccentric orbits so that they are capable of keeping more objects under their
dynamical control rather than passing them down to Saturn and Jupiter. It is well-known that Uranus and Neptune place objects into the
OC while Jupiter and Saturn tend to eject them (e.g. Duncan et al., 1987).\\

From our simulations we deduce that the population ratio between the OC and the SD is 12 $\pm$ 1. This ratio is a little lower than
that found by Dones et al. (2004) and Brasser et al. (2010), but similar to that of Kaib \& Quinn (2008). We attribute this to our
higher OC formation efficiency.\\

Having summarised the key properties of the OC we now turn to the SD and JFC production.

\subsection{Linking the SD and JFC populations}
Before addressing the population ratio between the SD and the JFC population we need to demonstrate that the SD produced in our
simulations is an acceptable source for the JFCs. In other words, we need to show that the the objects that evolve from the SD into the
JFC region reproduce the observed distribution of the actual JFCs acceptably well.\\

Here we use the method described in Levison \& Duncan (1997), who followed a large number of test particles from the Kuiper Belt into
the visible JFC region. Levison \& Duncan (1997) show that the inclination distribution of their simulated visible JFCs on their first
apparition is inconsistent with the observed one. Knowing that the typical inclination of the JFCs increases with time they use the
real distribution to determine the physical lifetime of the comets (also referred to as the fading time). They compare the inclination
distributions from their simulations with the observed one for various values of the physical age of the comets.\\

To do this, Levison \& Duncan (1997) define $\zeta(\tau,i)$ as the number of comets with inclinations between $i$ and $i+di$ and with
physical ages between $\tau$ and $\tau+d\tau$. Here $\tau$ is measured from the time when the comet first enters into the
visible ($q<2.5$~AU) region. By assuming that all comets fade instantaneously after a certain time $\tau_{\rm{f}}$ and remain dormant
afterwards, the inclination distribution of active comets is thus given by

\begin{equation}
 \xi_{\rm{a}}(i) = \int \limits_0^{\tau_{f}} \zeta(\tau,i) d\tau.
\end{equation}
We define similar distributions for the Tisserand parameter, $\xi_{\rm{a}}(T_J)$, and the minimum distance between Jupiter's orbit
and one of the comet's nodes, $\xi_{\rm{a}}(d_J)$. Here $d_J = {\rm min}(|a_J-r_\Omega|,|a_J-r_\mho|)$, where
$r_{\Omega,\mho}=a(1-e^2)/(1\mp e\cos \omega)$ are the distances of the comet's nodes to the Sun. Most JFCs are believed to be
scattered onto low-$q$ orbits by encounters with Jupiter and thus the distribution of $d_J$ should be indicative of the dynamical age
of these comets (Levison \& Duncan, 1997). Levison \& Duncan (1997) obtain $\xi_{\rm{a}}(i)$ and $\xi_{\rm{a}}(d_J)$ for a range of
$\tau_{\rm{f}}$ from their simulations. They compute the probability of a match between their simulations and the real comets by
performing a Komolgorov-Smirnov (KS) test for various values of $\tau_{\rm{f}}$. The peak of the KS probability as a function of the
active lifetime $\tau$ yields the best-fit value of $\tau_{\rm{f}}$. Their distribution peaks at $\tau_{\rm{f}} = 12$~kyr and they
conclude that this value must correspond to the typical fading time for the visible JFCs.\\

\begin{figure}
\resizebox{\hsize}{!}{\includegraphics[angle=-90]{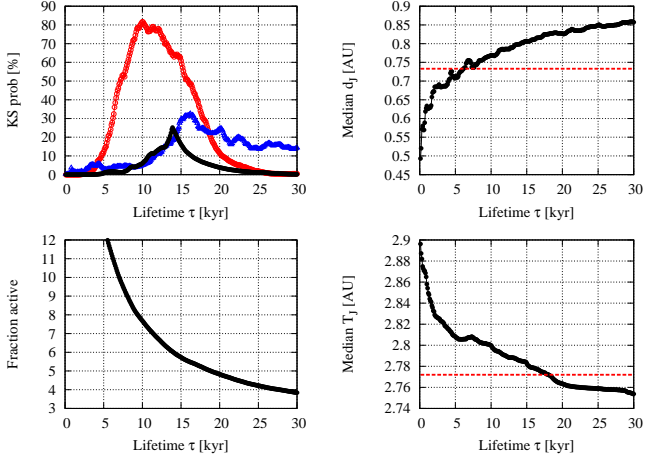}}
\caption{Top-left panel: Probability that the inclination (red), $d_J$ (blue) and $T_J$ (black) distributions of the visible JFCs
produced in our simulations matches their observed ones as a function of the ageing time, $\tau$. Top-right panel: the median value of
$d_J$ from our simulations as a function of ageing time. The red line indicates the observed value. Bottom-left: the ratio of the
number of dormant to active `visible' JFCs as a function of ageing time. Bottom-right: the median value of $T_J$. The red line
indicates the observed value.}
\label{tfade}
\end{figure}

We have repeated their procedure but included also the results from $\xi_{\rm{a}}(T_J)$. The outcome is shown in Fig.~\ref{tfade}. In
the top-right panel we depict the probability of the inclination (red), $d_J$ (blue) and $T_J$ (black) distributions matching their
observed distributions as a function of the physical lifetime, $\tau$. From the plot it appears that the best-fit value for the
physical lifetime, $\tau_{\rm{f}}$, lies between 10~kyr and 15~kyr, bracketing the value of 12~kyr found by Levison \& Duncan (1997).
Thus, we also adopt $\tau_{\rm{f}} = 12$~kyr here. Our match between the inclination and $d_J$ distribution is not as good as
Levison \& Duncan (1997) nor is the peak of the latter as high, but it is nevertheless a satisfactory match (KS probability larger than
20\%). The top-right panel depicts the median $d_J$ as a function of $\tau$, with the red line indicating the observed value. The
corresponding distribution for $T_J$ is shown in the bottom-right panel. Note that the median $d_J$ and $T_J$ distributions cross their
observed value at different $\tau$, indicating that the dynamics is not entirely controlled by scattering off of Jupiter. Last, the
bottom-left panel of Fig.~\ref{tfade} shows the ratio of the total number of JFCs (active and dormant) to the number of active ones.
The total number of comets with $\tau < \tau_{\rm{f}}$ is computed as $\eta(\tau_{\rm{f}})=\int \xi_{\rm{a}}(i) di$ and the total
number of JFCs is computed by taking $\tau \rightarrow \infty$.\\

For a nominal $\tau_{\rm{f}} = 12$~kyr, the total number of comets is about 6.5 times the number of active ones, similar to the factor
5 what was reported in Levison \& Duncan (1997) and Di Sisto et al. (2009). \\

\begin{figure}
\resizebox{\hsize}{!}{\includegraphics[angle=-90]{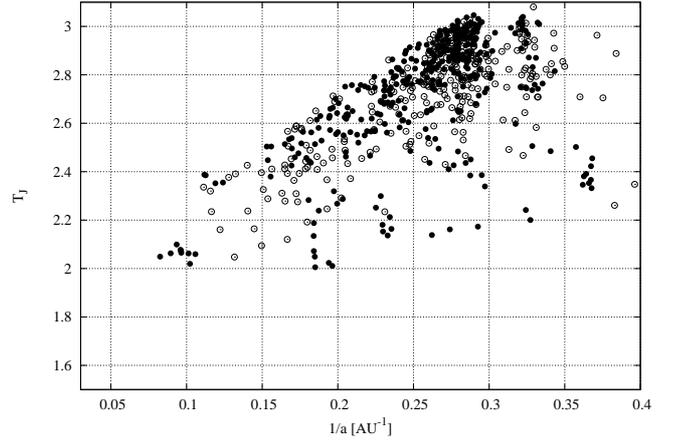}}
\caption{Plot of $1/a$ vs $T_J$ of the observed (open circles) and simulated comets (bullets) with $q<2.5$~AU and with $T_J<3.1$. The
data from the simulations are for the duration of the comets' active lifetime.}
\label{atj}
\end{figure}

There is an additional check we performed. We checked whether we generated a Saturn-family comets group i.e. comets that
were scattered into the inner Solar System by Saturn and have a node and/or their aphelion near Saturn. This would be problematic,
given that these comets are not observed. The Saturn-family comets would have $T_J<2.6$ and $a>6$~AU. In Fig.~\ref{atj} we have plotted
$T_J$ vs $1/a$ for all comets with $P<200$~yr, $q<2.5$~AU and $T_J<3.1$ (open circles). The observational data for this plot was
obtained from JPL\footnote{http://ssd.jpl.nasa.gov/sbdb\_query.cgi}. We superimposed the comets from our simulations during their
active lifetime (bullets). This lifetime is usually 12~kyr but shorter for some comets who were dynamically removed from the visible
region before having faded. As one may see the distribution of bullets with $(1/a,T_J)=(0.15,2.6)$ and lower are compatible with the
real comets. Thus we do not generate a swarm of Saturn-family comets.\\

We performed a few further tests, such as establishing the fraction of visible comets that are Halley-types, the average time spent
as a Centaur, the visible JFC perihelion distribution and the average dynamical lifetime of JFCs. The former two agree well with the
results of Levison \& Duncan (1997), but we find a shorter mean JFC dynamical lifetime: 165~kyr $\pm$ 60~kyr vs. 270~kyr for Levison \&
Duncan (1997).\\

In summary, we have a working model linking the SD population to the visible JFCs population and a reliable estimate for the physical
lifetime of comets.  We can now use this information to compute the SDO to JFC population ratio.\\

The formula relating the number of objects in the SD ($N_{\rm{SDO}}$) to the number of active objects in the JFC visible region
($N_{\rm{vJFC}}$) is:
\begin{equation}
 N_{\rm{SDO}} = \frac{N_{\rm{vJFC}}}{\tau_{\rm{vJFC}}|r_{\rm{SD}}|f_{\rm{vJFC}}},
\label{formula}
\end{equation}
where $r_{\rm{SD}}$ is the fractional decay rate of the SD at the current time, $f_{\rm{vJFC}}$ is the fraction of the comets escaping
from the SD that penetrate into the visible region and $\tau_{\rm{vJFC}}$ is the mean lifetime spent by these comets in the visible JFC
region as {\it active} comets. We evaluate these three quantities below.

\subsubsection{Evaluation of the fractional decay rate of the SD, $r_{\rm{SD}}$}
\begin{figure}
\resizebox{\hsize}{!}{\includegraphics[angle=-90]{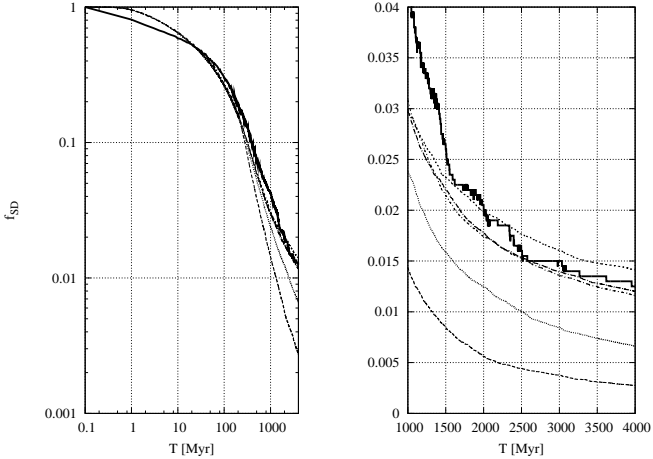}}
\caption{Remaining fraction of comets in the SD. The result of five simulations are shown. The solid line shows the remaining fraction
from from Duncan \& Levison (1997). Data kindly provided by Hal Levison. The dashed lines shows the results our simulations.}
\label{dsd}
\end{figure}

An SDO is considered to have left the SD if it has been removed from the simulation (through ejection or collision) or if it is still
evolving but has spent any time as a Centaur (i.e. has achieved $q<30$AU). Defining by $f_{\rm{SD}}(t)$ the fraction of the original
trans-Neptunian disk surviving in the SD at time $t$ (see Fig.~\ref{dsd}), the fractional decay rate of the SD population is
defined as $r_{\rm{SD}}(t)=(df_{\rm{SD}}(t)/dt)/f_{\rm{SD}}(t)$. We measured $r_{\rm{SD}}$ over the last 0.5~Gyr of our simulations
and obtained $\langle r_{\rm{SD}} \rangle =-(1.63 \pm 0.66)\times 10^{-10}$ per year. For reference, Duncan \& Levison (1997) report a
value of $\langle r_{\rm{SD}} \rangle =-2.7 \times 10^{-10}$ per year. This roughly factor of two difference is due to a higher
fraction of our SD residing in a fossilized state (resonant and detached objects) than Duncan \& Levison (1997). In fact, Gomes et al.
(2005b) and Gomes (2011) have shown that the migration of Neptune can create a ‘Fossilised’ Scattered Disc (FSD), which is comprised of
planetesimals that are either no longer interacting with Neptune, such as (136199) Eris, 2000 CR$_{105}$ and 2004 XR$_{190}$ (Buffy),
or planetesimals that are trapped in mean-motion resonances with Neptune for a long time.

\subsubsection{Evaluation of the fraction of JFCs, $f_{\rm{vJFC}}$}
To evaluate the fraction of SDOs that become visible JFCs precisely, we need to make sure we do not miss any object that enters the
$q<2.5$~AU region, even for a very short time. For this reason we have done an extra set of simulations where we remove objects when
they achieve this perihelion threshold. The check on the perihelion distance is done at every simulation timestep.\\

We find that of the comets that leave the SD, an average fraction $f_{\rm{vJFC}}=16.5\% \pm 8.0\%$ penetrate into the region of
visible JFCs. Our value of $f_{\rm{vJFC}}$ is somewhat lower than that reported in Levison and Duncan (1997) (30\%). We attribute the
difference to our source population having a different orbital distribution than that of Levison \& Duncan (1997).

\subsubsection {Evaluation of the active visible lifetime, $\tau_{\rm{vJFC}}$}
The average time any JFC spends with $q<2.5$~AU as an {\it active} object is computed using the simulations with a 100~yr
high-resolution output. More precisely, we assumed that at each output entry the comet spends the entire 100~yr output time in the
visible region. The comet is discarded once it reaches a physical lifetime of 12~kyr. We obtain $\tau_{\rm{vJFC}} = 2.6$~kyr, somewhat
shorter than the 3.6~kyr from Di Sisto et al. (2009). 

\subsubsection{The Scattered Disc to visible JFC population ratio}
To compute the total number of SDOs we assume that the distributions in $r_{\rm{SD}}$ and $f_{\rm{vJFC}}$ are Gaussian with means and
standard deviations given by the nominal and error values listed above. Plugging the values of all of the variables and their
distributions reported above into equation (\ref{formula}) we find a mean value of $N_{\rm{SDO}} = 1.5_{-0.6}^{+2.3} \times
10^7\,N_{\rm{vJFC}}$. The error-bars are once again 1-sigma\footnote{The 1-sigma error values were computed with Monte Carlo
simulations. We generated 10$^5$ values of $f_{\rm{vJFC}}$ and $r_{\rm{SD}}$ assumed they followed a Gaussian distribution with the
mean and the standard deviation reported above. From these we evaluated $N_{\rm{SD}}$. The 1-sigma uncertainty interval for
$N_{\rm{SD}}$ is given by the values that leave 15.8\% of the cumulative distribution of $N_{\rm{SD}}$ in each of the wings.} values.
By comparison, this relationship yields $N_{\rm{SDO}} = 6.0 \times 10^6\,N_{\rm{vJFC}}$ for Duncan \& Levison (1997). We use this new
SDO to JFC ratio in the next section to infer the OC/SD ratio from observations of size-limited populations of LPCs and JFCs.

\section{A re-examination of the Oort cloud to Scattered Disc population ratio}
In this section we attempt to obtain an updated value of the OC to SD population ratio for size-limited populations, rather than for
populations limited in total absolute magnitude. In other words we want to obtain a value for the number of Oort cloud objects,
$N_{\rm{OC}}$, and the number of SDOs, $N_{\rm{SD}}$, for comets of the same size. We use the flux of LPCs in the inner solar system as
a proxy for $N_{\rm{OC}}$ (Wiegert \& Tremaine, 1999) and the number of visible JFCs as a proxy for the number of SDOs (Duncan \&
Levison, 1997). The overall result depends on the flux of LPCs that enter the inner solar system and on the absolute magnitudes of an
LPC and a JFC of the same size. We discuss each of these below.

\subsection{The flux of new comets entering the inner solar system}
The flux of new comets entering the inner solar system is poorly known. The most-cited sources estimating the flux are Everhart
(1967), Hughes (2001) and Francis (2005). Most of these list the flux of new comets with a perihelion distance $q<4$~AU and absolute
magnitude $H_T < 10.9$. However, there is a potential problem with using the flux of new comets with $H_T < 10.9$: the sample may be
incomplete for new comets with $H_T > 6.5$. All three sources (Everhart, 1967; Hughes, 2001 and Francis, 2005) find a break in the
absolute magnitude distribution at $H_T \sim 6.5$. Indeed Francis (2005) finds that at higher absolute magnitudes the differential
absolute magnitude distribution is virtually flat i.e. the cumulative increases linearly with $H_T$ rather than as an exponential.
From their analysis of LPCs that come close to Earth Fern\'{a}ndez \& Sosa (2012) argue that the break in the absolute magnitude
distribution is caused by a corresponding change in the slope of the size-frequency distribution rather than observational
incompleteness. Thus in what follows we shall focus only on LPCs with $H_T < 6.5$.\\

Everhart (1967) finds that 8\,000 comets with $H_T < 10.9$ and $q<4$~AU should pass through perihelion in 127~yr. He uses a
photometric index $\nu=4$ so that for his data $H_T=H_{10}$. From his paper the flux ratio between comets with $H_T <6.5$ and $H_T <
10.5$ is 12, so that his flux for LPCs with $H_T<6.5$ is only 3.3~yr$^{-1}$ for $q<2.5$~AU. Approximately a third of LPCs are new
(Wiegert \& Tremaine, 1999; Fern\'{a}ndez \& Sosa, 2012), and thus Everhart's flux of new comets is 1.1~yr$^{-1}$ with $q < 2.5$~AU.
Hughes (2001) quotes an LPC flux of 0.53~yr$^{-1}$ per unit perihelion with $H_T<6.5$ and he also assumes a photometric index $\nu=4$.
When only a third of these are new his flux with $q<2.5$~AU becomes $0.44$~yr$^{-1}$. Francis (2005) determines the flux of new comets
to be 2.9~yr$^{-1}$ with $H_T < 10.9$ and $q<4$~AU. For new comets he uses a photometric index $\nu=2.44$ on the inbound leg, which is
Whipple's (1978) average for NCs. Approximately 40\% of his comets with $H_T < 10.9$ have $H_T < 6.5$ and thus his flux of new comets
is 0.73~yr$^{-1}$. Taking the average of the three sources, we arrive at an LPC flux of approximately (0.76 $\pm$ 0.33)~yr$^{-1}$ with
$q<2.5$~AU and $H_T<6.5$. We shall use this to determine $N_{\rm{OC}}$ below, but first we need to determine the size of an LPC with
$H_T<6.5$. This is done in the next subsection.

\subsection{The nuclear absolute magnitude of LPCs with $H_T=6.5$}
There have been a number of attempts to relate the cometary absolute magnitude to the absolute magnitude of the nucleus, $H$ (Bailey \&
Stagg, 1988; Weissman, 1996). However, these are generally unreliable. It is now well recognized that cometary nuclei develop
non-volatile, lag-deposit crusts that reduce the fraction of the nucleus surface available for sublimation (Brin \& Mendis, 1979;
Fanale \& Salvail, 1984). For most JFCs, the`active fraction', that is the fraction of the nucleus surface area that must be active to
explain the comet's water production rate, is typically only a few per cent, or even a fraction of a per cent (e.g. Fern\'{a}ndez et
al., 1999). For LPCs, however, the active fraction is very large, and can `exceed' 100\% (Sosa \& Fern\'{a}ndez, 2011). Thus LPCs are
brighter than JFCs of a comparable size at the same heliocentric distance. Given these higher activity levels, the discovery
probabilities for LPCs with small nucleus sizes should be considerably higher than those for JFCs of the same size. If the typical
JFC is larger than 2~km in diameter (the median value for 67 measured JFC nuclei is 3.7~km; Snodgrass et al., 2011), it is entirely
likely that an LPC with comparable brightness has a smaller, possibly sub-km nucleus. Is there some way of comparing values of
$H_T$ versus $H$?\\ 

Sosa \& Fern\'{a}ndez (2011) use observational data of the water production of LPCs and the subsequent non-gravitational forces to
derive a relation between the diameter of the nucleus of the LPC and its total absolute magnitude, which is given by

\begin{equation}
\log D = 1.2 - 0.13H_T,
\label{sf}
\end{equation}
where $D$ is given in kilometres. Equation (\ref{sf}) appears valid mostly for small comets (Fern\'{a}ndez \& Sosa, 2012). Equation
(\ref{sf}) should be compared with the usual equation relating diameter and absolute magnitude for non-active objects, for which $\log
D \propto 0.2H$. This difference in slope implies that for LPCs $H - H_T$ is not a constant, but depends on $H$ itself. 
\\

Substituting $H_T=6.5$ into equation (\ref{sf}) yields $D_{\rm{LPC}} = 2.3$~km, which corresponds to a nuclear magnitude of $H=17.3$
for a comet with a typical albedo of 4\% (Fern\'{a}ndez et al., 2001). However, the above relation does not hold for JFCs, because
these comets typically have a lower activity level than LPCs (e.g. Fern\'{a}ndez et al., 1999; Tancredi et al., 2006). To compare
apples to apples we need to know what is the value of $H_T$ for a JFC with $H=17.3$.\\

\subsection{The total absolute magnitude of JFCs with $H=17.3$}
Fern\'{a}ndez et al. (1999) plot the value of $H - H_T$ as a function of perihelion distance and fraction of active surface for JFCs
with radii between 1~km and 5~km. Limiting ourselves to JFCs with $q<2.5$~AU, we find $H_T \sim 9$ when $H\sim 17$ from their scatter
plots, and thus the typical difference is 8, give or take 1 magnitude. Fern\'{a}ndez \& Morbidelli (2006) also find $H - H_T  = 8 \pm
1$ from small, faint JFCs with $q<1.3$~AU. In what follows, we consider a JFC with $H = 17.3$ to have $H_T = 9.3$, with an error of
about 1 magnitudes. In conclusion, we find that LPCs with $H_T<6.5$ have the same nuclear magnitude (i.e. physical size) as JFCs
with $H_T<9.3$, give or take a magnitude. Thus, the OC to SD population ratio has to be determined from LPCs and JFCs with $H_T<6.5$
and $H_T< 9.3$ respectively, rather than from populations with the same limiting total absolute magnitude. We do this below.\\

\subsection{The Oort cloud to Scattered Disc population ratio}
To calibrate the number of SDOs with $H_T<9.3$ and subsequently compare this to the number of comets in the OC with the same diameter,
we need to know the number of visible JFCs with $H_T<9.3$. From their numerical simulations including physical ageing effects
Di Sisto et al. (2009) find that there are 117 active JFCs with $q<2.5$~AU and $D>2$~km. Levison \& Duncan (1997) estimate the number
of JFCs with $q<2.5$~AU and $H_T<9$ is 108, similar to that derived by Di Sisto et al. (2009).\\

The uncertainty in $H-H_T$ for the JFCs is of the order of 1 magnitude. The cumulative absolute magnitude distribution of the comets
obeys $N(>H) \propto 10^{-\alpha H}$. This corresponds to a cumulative size-frequency distribution of JFC nuclei $N(>D) \propto
D^{-\gamma}$ where $\gamma = 5\alpha$. For JFCs with diameters between approximately 2~km and 10~km the slope $\gamma \sim 2$ (e.g.
Lowry \& Weissman, 2003; Meech et al., 2004; Snodgrass et al., 2011), corresponding to $\alpha=0.4$. With this value of $\alpha$ the
JFC population varies by a factor of 2.5 for every magnitude difference between $H$ and $H_T$. Approximating the errors as Gaussian we
have $N_{\rm{vJFC}} = 117 \pm 50$ and the median total number of bodies in the SD with $D>2.3$~km is then $N_{\rm{SD}} =
1.7_{-0.9}^{+3.0} \times 10^9$. This is only a factor of three higher than Duncan \& Levison (1997).\\


We compute the number of comets in the OC with $H<17.3$ from the flux of new comets. As discussed earlier, this flux is about
(0.76 $\pm$ 0.33)~yr$^{-1}$ for LPCs with $H<17.3$ and $q<2.5$~AU. Kaib \& Quinn (2009) state that the average fraction of the OC that
has objects on orbits with $q<3$~AU is $10^{-11}$~yr$^{-1}$. Thus the total OC population for comets with $H<17.3$ is then $N_{\rm{OC}}
=(7.6 \pm 3.3) \times 10^{10}$. Once again assuming the OC population is normally distributed, the median population ratio between the
OC and SD for objects with $D>2.3$~km is then $44_{-34}^{+54}$. This is a factor of four higher than the results from our simulations
presented in the previous section. However, the error bar overlaps with the nominal value from the simulations.\\

What are the uncertainties in the above estimate? The above estimate of the OC/SD population ratio has taken into account the
uncertainties in the SD decay rate, visible JFC production and $H-H_T$ for JFCs and LPC flux. We did not yet consider other errors in
the OC population. \\

The most reliable estimates of the population of the OC yield of the order of 10$^{11}$ comets, but it is likely only to be correct
within a factor of two, or possibly even lower (Neslu\v{s}an, 2007). From the analysis of the motion of 26 LPCs and accounting for
non-gravitational forces Kr\'{o}likowska \& Dybczy\'{n}ski (2010) argue that approximately half of the LPCs that traditionally were
designated as being `new' may in fact already be `old'. This result would yield a revised population estimate of the OC ($4 \times
10^{10}$) that brings the OC to SD population ratio to $23_{-15}^{+26}$, the error bar once again overlapping with the nominal
value.\\

In summary, we find that our scenario of the contemporary formation of the OC and SD in the framework of the Nice predicts a OC/SD
ratio that is about four times lower than the ratio deduced from observations. The latter, however, has large uncertainties, but the
model-predicted and observation-deduced values do agree within the error bars. Therefore, the agreement between the simulations
and observations cannot be rejected by a null hypothesis and thus there might be no problem with our scenario of the formation of
these two comet reservoirs.

\section{Conclusions}
We have performed simulations of the formation and evolution of the OC and SD in the framework of the Nice model. For OC formation
simulations we kept the Sun in the current Galactic environment. The simulations lead to a somewhat higher capture efficiency than
those of the more classical model where the giant planets are assumed to be on current orbits (e.g. Dones et al., 2004; Kaib \& Quinn,
2008; Dybczy\'{n}ski et al., 2008; Brasser et al., 2010). We find that the efficiency of trapping comets in the OC is $\sim 7\%$ and
that the simulated OC to SD population ratio is approximately 12 $\pm$ 1, somewhat lower than most earlier results but still of the
same order. \\

We have shown that the SD produced in our simulations generates a population of JFCs that is consistent with the observed population.
This is the first time that a dynamically hot SD is shown to be consistent with the JFC population. A previous model of JFC origin
started from a dynamically cold SD (Duncan \& Levison, 1997), whose existence is challenged by observations.\\

Using the link that we have established between the SD and the JFC population, as well as the link between the OC and the LPC
population described in Wiegert \& Tremaine (1999), we deduced a OC/SD population ratio from the observed fluxes of LPCs and JFCs. We
performed the calculations for size-limited samples of comets (i.e. for LPCs and JFCs with the same diameter of the nucleus), by using
the most recent conversions from total magnitude to nuclear magnitude available in the literature for LPCs and JFCs. We found a
population ratio of $N_{\rm{rat}} = 44_{-34}^{+54}$, roughly a factor of four higher than the simulated nominal value of 12. The error
bar of the observed ratio overlaps with the nominal value from simulations. This result takes into account all known uncertainties.
Thus we conclude that our scenario of contemporary formation of the OC and SD in the framework of the Nice model is not inconsistent
with the observations.\\

Our scenario has several implications. First, given that on average the current SD population retained just 0.95\% of the
original trans-Neptunian disc population, we can estimate that the latter contained $1.9\times 10^{11}$ comets with $H<17.3$
($D>2.3$~km) at the time of the instability of the giant planets. If instead we use the current OC population for this estimate, we
find $10^{12}$ comets. For comparison, the model by Morbidelli et al. (2009) of the primordial trans-Neptunian disc predicted 5$\times
10^{11}$ comets for the same value of $H$.\\

Second, in our scenario both the OC and the SD are derived from the {\it same} parent population, i.e. the primordial trans-Neptunian
disc. Thus, the LPCs and JFCs that come from these reservoirs should share (on average) the same physical properties. This means that
they should have the same size distribution and the same range of chemical compositions. Both the size distributions and chemical
compositions are very uncertain at the current stage of observational art. However, we remark that the size distribution of LPCs
brighter than $H_T=6.5$ and that of JFCs brighter than $H_T=9$ are fairly similar; both are compatible with a cumulative size
distribution with a slope of $\gamma=-2$ (Lowry \& Weissman, 2003; Meech et al., 2004; Snodgrass et al., 2011; Fern\'{a}ndez \& Sosa,
2012). We also remark that the slope of the size distribution is very different for fainter JFCs, but this may be due to observational
incompleteness or break-up of the comets on their way into the inner solar system. Regarding the chemical compositions, A'Hearn et al.
(2012) argue that, from a statistical point of view, LPCs and JFCs are indistinguishable. Thus, although the last word has still to be
said from the observational viewpoint, the prediction provided by our scenario seems to be verified.

\section{Acknowledgements}
\footnotesize{We are indebted to Paul Weissman for valuable discussions during an earlier version of this manuscript and for some
absolute magnitude data of JFCs. We also thank Hal Levison and Martin Duncan for stimulating discussions. We are grateful to Hal
Levison and Julio Fer\'{a}ndez who acted as reviewers and their criticisms greatly improved the quality of this paper. AM gratefully
acknowledges financial support from Germany's Helmholtz Alliance through their `Planetary Evolution and Life' programme. The Condor
Software Program (Condor) was developed by the Condor Team at the Computer Sciences Department of the University of Wisconsin-Madison.
All rights, title, and interest in Condor are owned by the Condor Team.\\}

\section{Bibliography}
{\footnotesize
A'Hearn, M.~F., and 22 colleagues 2012.\ Cometary Volatiles and the Origin of Comets.\ The Astrophysical Journal 758, 29. \\
Bailey, M.~E., Stagg, C.~R.\ 1988.\ Cratering constraints on the inner Oort cloud - Steady-state 
models.\ Monthly Notices of the Royal Astronomical Society 235, 1-32. \\
Brasser, R., Duncan, M.~J., Levison, H.~F.\ 2006.\ Embedded star clusters and the formation of the Oort Cloud.\ Icarus 184, 59-82. \\
Brasser, R., Duncan, M.~J., Levison, H.~F.\ 2007.\ Embedded star clusters and the formation of the Oort cloud. II. The effect of the
primordial solar nebula.\ Icarus 191, 413-433. \\
Brasser, R., Morbidelli, A., Gomes, R., Tsiganis, K., Levison, H.~F.\ 2009.\ Constructing the secular architecture of the solar system
II: the terrestrial planets.\ Astronomy and Astrophysics 507, 1053-1065. \\
Brasser, R., Higuchi, A., Kaib, N.\ 2010.\ Oort cloud formation at various Galactic distances.\ Astronomy and Astrophysics 516, A72. \\
Brasser, R., Duncan, M.~J., Levison, H.~F., Schwamb, M.~E., Brown, M.~E.\ 2012.\ Reassessing the formation of the inner Oort cloud in
an embedded star cluster.\ Icarus 217, 1-19.\\
Brin, G.~D., Mendis, D.~A.\ 1979.\ Dust release and mantle development in comets.\ The Astrophysical Journal 229, 402-408. \\
Chambers, J.~E.\ 1999.\ A hybrid symplectic integrator that permits close encounters between massive bodies.\ Monthly Notices of the
Royal Astronomical Society 304, 793-799. \\
Dones, L., Weissman, P. R., Levison, H. F., Duncan, M. J. 2004. Oort cloud formation and dynamics. In: Comets II, M. C. Festou, H. U.
Keller, and H. A. Weaver (eds.), University of Arizona Press, Tucson, AZ, 153-174.\\
Duncan, M., Quinn, T., Tremaine, S.\ 1987.\ The formation and extent of the solar system comet cloud.\ The Astronomical Journal 94,
1330-1338. \\
Duncan, M.~J., Levison, H.~F.\ 1997.\ A scattered comet disk and the origin of Jupiter family comets.\ Science 276, 1670-1672. \\
Dybczy{\'n}ski, P.~A., Leto, G., Jakub{\'{\i}}k, M., Paulech, T., Neslu{\v s}an, L.\ 2008.\ The simulation of the outer Oort cloud
formation. The first giga-year of the evolution.\ Astronomy and Astrophysics 487, 345-355. \\
Everhart, E.\ 1967.\ Intrinsic distributions of cometary perihelia and magnitudes.\ The Astronomical Journal 72, 1002. \\
Fanale, F.~P., Salvail, J.~R.\ 1984.\ An idealized short-period comet model - Surface insolation, H$_2$O flux, dust flux, and mantle
evolution.\ Icarus 60, 476-511. \\
Fern{\'a}ndez, J.~A., Tancredi, G., Rickman, H., Licandro, J.\ 1999.\ The population, magnitudes, and sizes of Jupiter family comets.\
Astronomy and Astrophysics 352, 327-340.\\
Fern{\'a}ndez, J.~A., Brunini, A.\ 2000.\ The buildup of a tightly bound comet cloud around an early Sun immersed in a dense Galactic
environment: Numerical experiments.\ Icarus 145, 580-590. \\
Fern{\'a}ndez, J.~A., Morbidelli, A.\ 2006.\ The population of faint Jupiter family comets near the Earth.\ Icarus 185, 211-222.\\
Fern{\'a}ndez, J.~A., Sosa, A.\ 2012.\ Magnitude and size distribution of long-period comets in Earth-crossing or approaching orbits.\
Monthly Notices of the Royal Astronomical Society 423, 1674-1690.\\
Fern{\'a}ndez, Y.~R., Jewitt, D.~C., Sheppard, S.~S.\ 2001.\ Low Albedos Among Extinct Comet Candidates.\ The Astrophysical Journal
553, L197-L200. \\
Francis, P.~J.\ 2005.\ The Demographics of Long-Period Comets.\ The Astrophysical Journal 635, 1348-1361. \\
Gladman, B., Marsden, B.~G., Vanlaerhoven, C.\ 2008.\ Nomenclature in the Outer Solar System.\ The Solar System Beyond Neptune 43-57.
Edited by M. A. Barucci; H. Boehnhardt; D. P. Cruikshank; A. Morbidelli; R. Dotson. University of Arizona press, Tucson, AZ, USA\\
Gomes, R., Levison, H.~F., Tsiganis, K., Morbidelli, A.\ 2005a.\ Origin of the cataclysmic Late Heavy Bombardment period of the
terrestrial planets.\ Nature 435, 466-469. \\
Gomes, R.~S., Gallardo, T., Fern{\'a}ndez, J.~A., Brunini, A.\ 2005b.\ On The Origin of The High-Perihelion Scattered Disk: The Role of
The Kozai Mechanism And Mean Motion Resonances.\ Celestial Mechanics and Dynamical Astronomy 91, 109-129. \\
Gomes, R.~S.\ 2011.\ The origin of TNO 2004 XR$_{190}$ as a primordial scattered object.\ Icarus 215, 661-668.\\
Heisler, J., Tremaine, S.\ 1986.\ The influence of the galactic tidal field on the Oort comet cloud.\ Icarus 65, 13-26. \\
Hills, J.~G.\ 1981.\ Comet showers and the steady-state infall of comets from the Oort cloud.\ The Astronomical Journal 86, 1730-1740.
\\
Holmberg, J., Flynn, C.\ 2000.\ The local density of matter mapped by Hipparcos.\ Monthly Notices of the Royal Astronomical Society
313, 209-216. \\
Hughes, D.~W.\ 2001.\ The magnitude distribution, perihelion distribution and flux of long-period comets.\ Monthly Notices of the Royal
Astronomical Society 326, 515-523.\\
Kaib, N.~A., Quinn, T.\ 2008.\ The formation of the Oort cloud in open cluster environments.\ Icarus 197, 221-238. \\
Kaib, N.~A., Quinn, T.\ 2009.\ Reassessing the Source of Long-Period Comets.\ Science 325, 1234. \\
Kaib, N.~A., Quinn, T., Brasser, R. \ 2011. \ Decreasing Computing Time with Symplectic Correctors in Adaptive Timestepping Routines.
The Astronomical Journal 141, 1-7. \\
Kr{\'o}likowska, M., Dybczy{\'n}ski, P.~A.\ 2010.\ Where do long-period comets come from? 26 comets from the non-gravitational Oort
spike.\ Monthly Notices of the Royal Astronomical Society 404, 1886-1902. \\
Leto, G., Jakub{\'{\i}}k, M., Paulech, T., Neslu{\v s}an, L., Dybczy{\'n}ski, P.~A.\ 2009.\ 2-Gyr Simulation of the Oort-cloud
Formation II. A Close View of the Inner Oort cloud after the First Two Giga-years.\ Earth Moon and Planets 105, 263-266. \\
Levison, H.~F., Duncan, M.~J., 1994. The long-term dynamical behavior of shortperiod comets. Icarus 108, 18-36.\\
Levison, H.~F.\ 1996.\ Comet Taxonomy.\ Completing the Inventory of the Solar System 107, 173-191.\\
Levison, H.~F., Duncan, M.~J.\ 1997.\ From the Kuiper Belt to Jupiter-Family Comets: The Spatial Distribution of Ecliptic Comets.\
Icarus 127, 13-32. \\
Levison, H.~F., Dones, L., Duncan, M.~J.\ 2001.\ The Origin of Halley-Type Comets: Probing the Inner Oort Cloud.\ The Astronomical
Journal 121, 2253-2267. \\
Levison, H.~F., Duncan, M.~J., Dones, L., Gladman, B.~J.\ 2006.\ The scattered disk as a source of Halley-type comets.\ Icarus 184,
619-633. \\
Levison, H.~F., Morbidelli, A., Vokrouhlick{\'y}, D., Bottke, W.~F.\ 2008a.\ On a Scattered-Disk Origin for the 2003 EL$_{61}$
Collisional Family--An Example of the Importance of Collisions on the Dynamics of Small Bodies.\ The Astronomical Journal 136,
1079-1088. \\
Levison, H.~F., Morbidelli, A., Vanlaerhoven, C., Gomes, R., Tsiganis, K.\ 2008b.\ Origin of the structure of the Kuiper belt during a
dynamical instability in the orbits of Uranus and Neptune.\ Icarus 196, 258-273. \\
Levison, H.~F., Duncan, M.~J., Brasser, R., Kaufmann, D.~E.\ 2010.\ Capture of the Sun's Oort Cloud from Stars in Its Birth Cluster.\
Science 329, 187. \\
Levison, H.~F., Morbidelli, A., Tsiganis, K., Nesvorn{\'y}, D., Gomes, R.\ 2011.\ Late Orbital Instabilities in the Outer Planets
Induced by Interaction with a Self-gravitating Planetesimal Disk.\ The Astronomical Journal 142, 152. \\
Lowry, S.~C., Weissman, P.~R.\ 2003.\ CCD observations of distant comets from Palomar and Steward Observatories.\ Icarus 164,
492-503.\\
MacMillan, P.~J., Binney, J.~J.\ 2010.\ The uncertainty in Galactic parameters.\ Monthly Notices of the Royal Astronomical Society 402,
934-940. \\
Meech, K.~J., Hainaut, O.~R., Marsden, B.~G.\ 2004.\ Comet nucleus size distributions from HST and Keck telescopes.\ Icarus 170,
463-491.\\
Morbidelli, A., Levison, H.~F., Tsiganis, K., Gomes, R.\ 2005.\ Chaotic capture of Jupiter's Trojan asteroids in the early Solar
System.\ Nature 435, 462-465. \\
Morbidelli, A., Tsiganis, K., Crida, A., Levison, H.~F., Gomes, R.\ 2007.\ Dynamics of the Giant Planets of the Solar System in the
Gaseous Protoplanetary Disk and Their Relationship to the Current Orbital Architecture.\ The Astronomical Journal 134, 1790-1798. \\
Morbidelli, A., Levison, H.~F., Bottke, W.~F., Dones, L., Nesvorn{\'y}, D.\ 2009.\ Considerations on the magnitude distributions of
the Kuiper belt and of the Jupiter Trojans.\ Icarus 202, 310-315. \\
Morbidelli, A., Brasser, R, Gomes, R., Levison, H, Tsiganis, K., 2010. Evidence from the asteroid belt for a violent evolution of
Jupiter's orbit.\ The Astronomical Journal, 140, 1-11.\\
Neslu\v{s}an, L.\ 2007.\ The fading problem and the population of the Oort cloud.\ Astronomy and Astrophysics 461, 741-750. \\
Neslu{\v s}an, L., Dybczy{\'n}ski, P.~A., Leto, G., Jakub{\'{\i}}k, M., Paulech, T.\ 2009.\ 2-Gyr Simulation of the Oort-Cloud
Formation. I.
Introduction on a New Model of the Outer Oort-Cloud Formation.\ Earth Moon and Planets 105, 257-261. \\
Oort, J.~H.\ 1950.\ The structure of the cloud of comets surrounding the Solar System and a hypothesis concerning its origin.\ Bulletin
of the Astronomical Institutes of the Netherlands 11, 91-110. \\ 
Petit, J.-M., Morbidelli, A., Chambers, J.\ 2001.\ The Primordial Excitation and Clearing of the Asteroid Belt.\ Icarus 153, 338-347.\\
Rickman, H., Fouchard, M., Froeschl{\'e}, C., Valsecchi, G.~B.\ 2008.\ Injection of Oort Cloud comets: the fundamental role of stellar
perturbations.\ Celestial Mechanics and Dynamical Astronomy 102, 111-132. \\
Di Sisto, R.~P., Fern{\'a}ndez, J.~A., Brunini, A.\ 2009.\ On the population, physical decay and orbital distribution of Jupiter family
comets: Numerical simulations.\ Icarus 203, 140-154. \\
Snodgrass, C., Fitzsimmons, A., Lowry, S.~C., Weissman, P.\ 2011.\ The size distribution of Jupiter Family comet nuclei.\ Monthly
Notices of the Royal Astronomical Society 414, 458-469.\\
Sosa, A., Fern{\'a}ndez, J.~A.\ 2011.\ Masses of long-period comets derived from non-gravitational effects - analysis of the computed
results and the consistency and reliability of the non-gravitational parameters.\ Monthly Notices of the Royal Astronomical Society
416, 767-782.\\
Tancredi, G., Fern{\'a}ndez, J.~A., Rickman, H., Licandro, J.\ 2006.\ Nuclear magnitudes and the size distribution of Jupiter family
comets.\ Icarus 182, 527-549. \\
Tsiganis, K., Gomes, R., Morbidelli, A., Levison, H.~F.\ 2005.\ Origin of the orbital architecture of the giant planets of the Solar
System.\ Nature 435, 459-461. \\
Weissman, P.~R.\ 1980.\ Physical loss of long-period comets.\ Astronomy and Astrophysics 85, 191-196. \\
Weissman, P.~R.\ 1983.\ The mass of the Oort cloud.\ Astronomy and Astrophysics 118, 90-94. \\
Weissman, P.~R.\ 1996.\ The Oort Cloud.\ Completing the Inventory of the Solar System, Astronomical Society of the Pacific Conference
Proceedings, volume 107, T.W. Rettig and J.M. Hahn, Eds., 265-288\\
Whipple, F.~L.\ 1978.\ Cometary brightness variation and nucleus structure.\ Moon and Planets 18, 343-359. \\
Wiegert, P., Tremaine, S.\ 1999.\ The Evolution of Long-Period Comets.\ Icarus 137, 84-121.\\
Wisdom, J., Holman, M., Touma, J.\ 1996.\ Symplectic Correctors.\ Fields Institute Communications, 
Vol.~10, 217-227. }
\end{document}